\documentclass[preprint,pre,aps,superscriptaddress,a4paper]{revtex4}
\usepackage[dvips]{graphicx}

\usepackage[utf8]{inputenc}
\usepackage[T1]{fontenc}
\usepackage{amsmath}
\usepackage{float}

\usepackage{epsfig,amsmath,amsfonts,amssymb,
color,
}
\usepackage{siunitx}
\usepackage{comment}

\begin{document}

\title{
Shear viscosity of pseudo hard-spheres
}

\author{Faezeh Pousaneh$^{\scriptscriptstyle{}}$\footnote{Corresponding
    author. Electronic address: \texttt{faezeh.pousaneh@ntnu.no}.}}  
%
%
\author{Astrid S.\ de Wijn  }
\affiliation{Department of Mechanical and Industrial Engineering, Norwegian University of
Science and Technology, 7491 Trondheim, Norway}

\begin{abstract}
We present molecular dynamics simulations of pseudo hard sphere fluid (generalized WCA  potential with exponents (50, 49) proposed by Jover {\it {et al.}} {\it J.~Chem.~Phys} {\bf 137}, (2012))  using GROMACS package.  The equation of state and radial distribution functions at contact are obtained from simulations and compared to the available theory of true hard spheres (HS) and available data on pseudo hard spheres.  The comparison shows agreements with  data by  Jover {\it {et al.}} and the  Carnahan-Starling equation of HS. The shear viscosity is obtained from the simulations and  compared to the Enskog expression and previous HS simulations. It is demonstrated that the PHS potential  reproduces the HS shear viscosity accurately. 

\end{abstract}
\maketitle

\section{Introduction}
Hard Sphere (HS) models have been widely used as a basic approximation of a spherical atom or molecule (See, for example ~\cite{pippo}), because of the simplicity of the interaction potential and the instantaneous elastic collision dynamics.
Although the HS model is an idealized model, it still captures the essential physics of macroscopic behavior of real fluids, both in and out of equilibrium.
Consequenly, HS-based models are often used to study and understand the thermodynamics and transport properties of liquids.
Nevertheless, because HS models are highly idealised, it is difficult to make quantitative predictions for more complex molecules based on purely theoretical calculations.
This is why theory based hard sphere models are often used as a basis for empirical approaches to fluid properties that go far beyond spherical molecules~\cite{VW,viscvogelmethane}.
This type of approach, however, requires analytical theory that can be challenging to develop, especially for more complicated models.

Because of the simple instantaneous dynamics, there are a great deal of analytical results for HS models for properties of liquids, such as equations of state~\cite{carnahanstarling} and transport coefficients~\cite{pippo,Chapman:52:0}.
These kinds of results can still be challenging to obtain, but they provide a powerful basis for continuing development of fluid theory~\cite{safths}.
There are also a number of extensions of HS model that still retain the instantaneous collision dynamics and therefore are still somewhat tractable when it comes to analytical approaches.
Examples are sperocylinders~\cite{spherocylinders}, rough hard spheres~\cite{Chapman:52:0}, loaded hard spheres~\cite{loadedspheres}, and dipolar hard spheres~\cite{dhsshear,dhsrelaxation}.

In supporting this development molecular dynamics (MD) simulations are tools that have become much more commonplace, for example as numerical experiments to verify the theoretical results.
Although there are available MD studies on transport coefficients of HS fluids, {  for example a comprehensive one by Sigurgeirsson  {\it {et al.}} ~\cite{heyes:09:0},} it is not possible to directly use the state of art simulations packages like LAMMPS, GROMACS or  NAMD  which provide high-speed simulations of complex and large systems  \cite{Plimpton:95:0,Spoel:05:0,Phillips:05:0}.
These simulation packages rely on approximately smooth dynamics, and thus do not support the HS model's instantaneous collision dynamics.

To get around this, 
Jover {\it {et al.}} proposed a pseudo hard sphere (PHS) model
which
is a cut and shifted version of a Mie potential with exponents
(50,49) \cite{jover:12:0}.
Using this nearly hard, but smooth potential, Jover {\it {et al.}}\ have been able to reproduce
structural and thermodynamic properties  accurately 
compared with available simulation data for the original
HS system.
It has also been shown to provide reliable results for
studying liquid-solid coexistence~\cite{vega:13:0}. 

The focus of the PHS research mentioned above has been primarily on equilibrium properties.
However, non-equilibrium properties such as transport coefficients are much more sensitive to changes in the dynamics than equilibrium properties.
Because of this, a good reproduction of equilibrium properties does not directly imply that non-equilibrium properties would be described as well.
So far, only the self-diffusion was briefly tested in~\cite{jover:12:0}.
The performance of the PHS potential for non-equilibrium properties has therefore not been sufficiently established.

In this work, we test in detail the reliability of the (50,49) PHS potential for calculating the viscosity.
{ We investigate the density dependence of the viscosity, and compare it to both analytical results for HS and previous MD simulations of true HS~\cite{heyes:09:0}.  }
\section{Simulation setup}
{ The Mie ($\lambda_r, \lambda_a$) potential (the generalized Lennard-Jones) can be changed to a repulsive potential by shifting it by its well-depth value and equating the larger distances interactions to zero. WCA potential is one of this cut-and-shift potential which has the  form 
\[ U_{WCA}(r)=
  \begin{cases} 
   4 \epsilon \big[ \big(\frac{\sigma}{r}\big)^{12}-\big(\frac{\sigma}{r}\big)^{6 }  \big]+\epsilon; \hspace{2cm}  r<   2^{1/6} \sigma \\
   \hspace{3.5cm}  0  ; \hspace{2cm}     r \geq   2^{1/6}\sigma
   \label{eqWCA}
  \end{cases}.
\]
\vspace{-1cm}
\begin{equation}
\end{equation}
The WCA potential can be used to present a HS system in an approximate manner. However, the reliability of the approximation depends on  ($\lambda_r, \lambda_a$) and on the temperature of the system.  Jover {\it {et al.}}~\cite{jover:12:0} have  studied the effect of  exponents  $\lambda_r$ and $\lambda_a$ on the steepness and the shape of the potential. They have chosen the pair (50,49) as a compromise between fidelity of the representation of the HS and the size of the time step in MD simulations.  The steeper the potential the greater the fidelity, but the smaller the time step that is required. 
The PHS potential  proposed by  Jover {\it {et al.}}~\cite{jover:12:0}  has the following form
\[U_{50,49}(r)=
  \begin{cases} 
   50 ~ (\frac{50}{49})^{49} \epsilon \big[ \big(\frac{\sigma}{r}\big)^{50 }-\big(\frac{\sigma}{r}\big)^{49 }  \big]+\epsilon; \hspace{2cm}  r<   \frac{50}{49} \sigma \\
   \hspace{5.cm}  0  ; \hspace{2cm}     r \geq   \frac{50}{49} \sigma
   \label{eqPHS}
  \end{cases}.
\]
\vspace{-1cm}
\begin{equation}
\end{equation}
In addition the effect of  temperature on $U_{50,49}(r)$ potential is studied in ref.~\cite{jover:12:0} and is  concluded that at  a reduced temperature of $ T^*=\frac{\epsilon}{k_B T}=2/3$ the potential  produces better agreement with  the true HS.   They have  examined  this  in terms of thermodynamics and structural properties.}

\label{sec3}
In order to simulate the PHS model, we have used GROMACS  version 5. The  potential is implemented as a  tabular form. The first goal was to obtain the equation of state (pressure versus density). We have simulated a box of  $N=1000$  particles  with  LJ parameters  $\sigma$, $\epsilon$ at different pressures  corresponding to different densities, at reduced temperature   $ T^*=2/3$. We limit ourselves to this  temperature because according to Refs.~\cite{vega:13:0,jover:12:0} the equilibrium properties of the PHS and HS models are similar at this temperature. In what follows, all quantities are given in reduced units: $t^*=t\sqrt{ \frac{k_B T}{\sigma^2 m}}$, $r^*= \frac{r}{\sigma}$, $\rho^*= \rho \sigma^3$ and $P^*=  \frac{P \sigma^3}{k_B T}$,  $\eta^*=\eta (\frac{\sigma ^2 }{\sqrt {mk_B T}})$, where $\rho$, $P$ and $\eta$ denote the number density, the pressure, and the viscosity respectively. 
 The equilibrated systems are simulated  in NPT ensemble for  $t^*=20000 $ with time-step of $\delta t^*=0.0011$ using a Parrinello-Rahman barostat and a velocity-rescale thermostat.

\begin{figure}
\centering
\hspace{-0cm}
\includegraphics[scale=0.6]{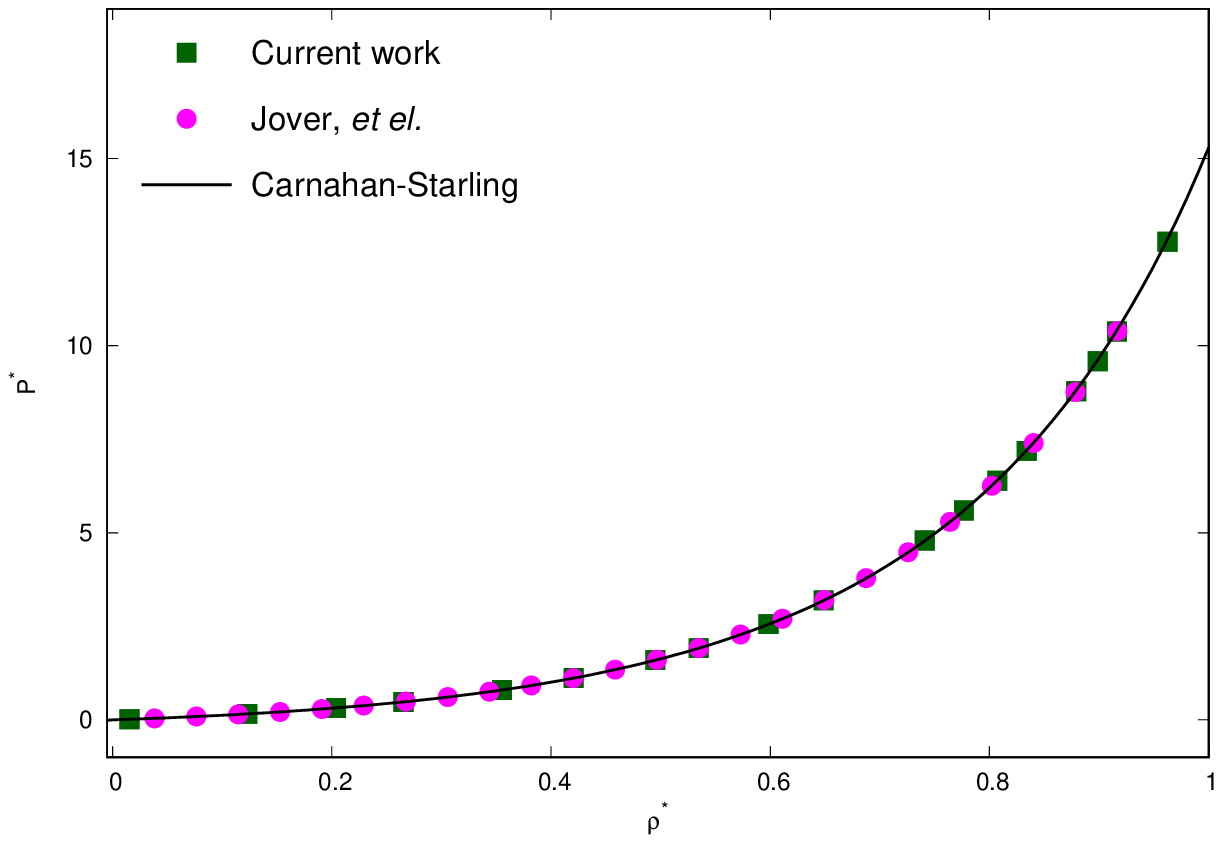}
\includegraphics[scale=0.6]{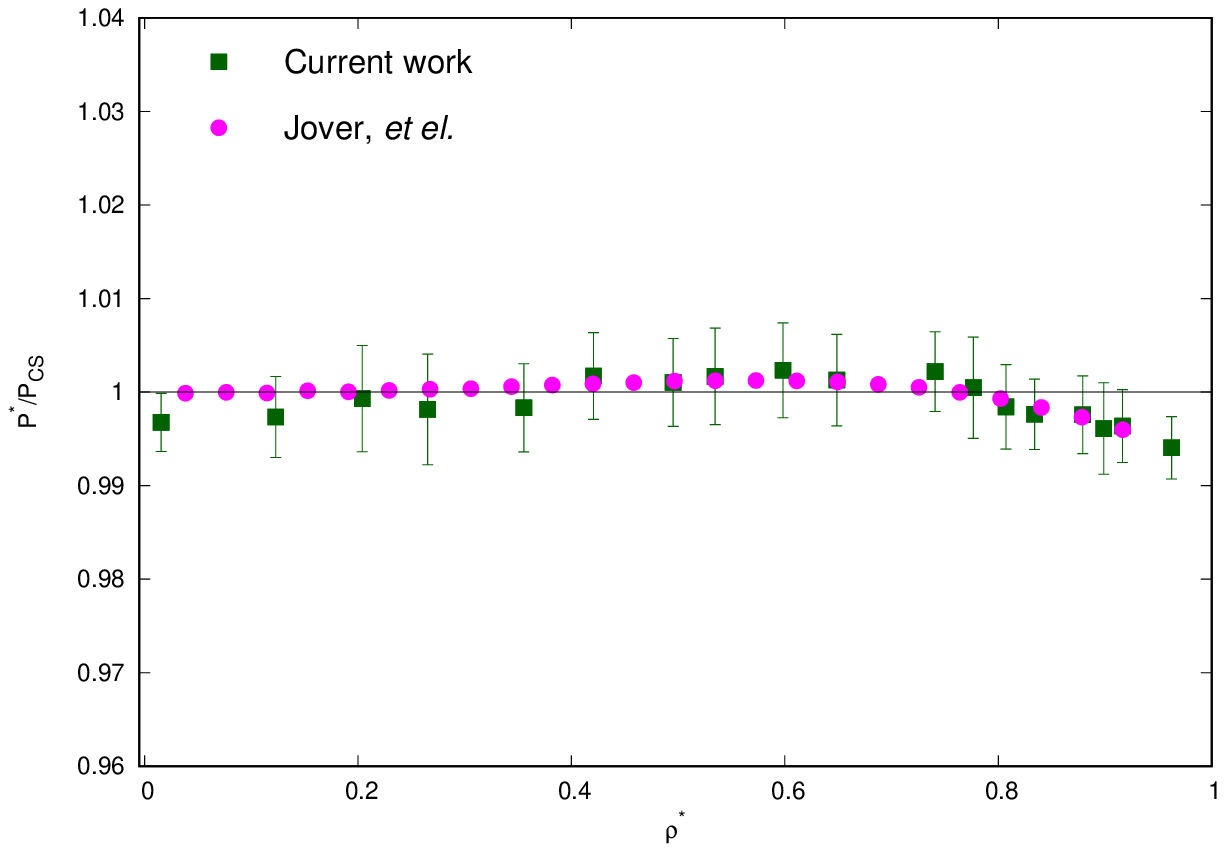}
\caption{(Left) Reduced pressure versus reduced number density from MD simulation of PHS in this work (squares) is compared to earlier work by Jover {\it {et al.}} ~\cite{jover:12:0} (circles) and theoretical Carnahan-Starling equation of state (solid line).{  (Right) The same as left plot, but rescaled with the Carnahan-Starling pressure, in order to enhance details. } The errors in  the current work data of $P^*$ are less than 4\%, so are not well visible.
}
\label{PHS}
\end{figure}

 Simulation results for the equation of state are  shown in Fig.~\ref{PHS}.  The relative errors in the pressure  $P^*$  are less than 4\%. The results agree with the results by Jover {\it {et al.}}~\cite{jover:12:0}. 
They are also in good agreement with the Carnahan-Starling equation of state (solid line) \cite{Carnahan:69:00} 
\begin{align} 
P^*= (6/\pi)\frac{\xi(1+\xi^*+\xi^{*2}-\xi^{*3})}{(1-\xi^*)^3}, \\
 \xi^*=(\pi/6) \rho^*.\hspace{1.5cm}
\end{align} 
Ref.~\cite{vega:13:0} also reports similar results using GROMACS package. 

\subsection{Shear viscosity} 
The shear viscosity can be determined both by equilibrium molecular-dynamics (EMD) or non-equilibrium molecular-dynamics (NEMD) simulations.
EMD methods are based on pressure or momentum fluctuations, for example the Transverse-Current Autocorrelation Function  (TCAF) method or Green-kubo method.
The TCAF method is the easiest to implement and has several  advantages over the Green-kubo relation. The first advantage is that any non-hydrodynamic behavior is easy to identify in the TCAF. The second advantage of the TCAF method is that it provides a natural way to estimate the magnitude of finite-size effects.
This can
be done in a single simulation and the results extrapolated to the infinite system limit in a straightforward calculation, (see \cite{palmer:94:0} and reference therein).
In NEMD methods such as  periodic perturbation (PP) method, instead of measuring intrinsic fluctuations   an  external force is applied to the system.  The magnitude of this force is chosen such that the effects are much easier to detect  than the internal fluctuations \cite{zhao:08:00,hess:02:0,Gregori:12:0,Gaskell:74:00}. 

There are many works studying shear viscosity  using MD simulations, both EMD and NEMD methods \cite{zhao:08:00,Gaskell:74:00,palmer:94:0,plathe:02:00,sunda:13:00,Gezeltr:10:0,Rowley:07:0,Sun:08:0,hess:02:0}.
Here we follow the work by B. Hess \cite{hess:02:0} which studies shear viscosity determination using GROMACS. We obtain shear viscosities of PHS model from  two methods:  TCAF and PP. We explain both methods shortly below.

\subsubsection{Transverse-current autocorrelation function}
Consider an incompressible liquid with an initial velocity field generated for example by equilibrium thermal fluctuations.
The velocity field can be decomposed into plane waves of the form ${\bold u} (x,0,0)=u_0 \cos(kz)$.
The solution to the Navier-Stokes equation for these components is then of the form
\begin{equation}
u_x (z,t) = u_0 e^{-t/\tau_r} \cos (kz); \hspace{2cm} \tau_r=\frac{\rho}{\eta k^2}.
\label{NS}
\end{equation} 
The solution indicates that the plane waves decay exponentially with a time constant which is inversely proportional to the viscosity $\eta$. 
However, at microscopic level and short time scales the behavior is not purely exponential. To account for this, a phenomenological correction can be applied by incorporating a relaxation time,  leading to different solution to Eq.~\ref{NS} (for details see ref.~\cite{hess:02:0});
\begin{equation}
u_x (z,t) = u_0 e^{-t/(2\tau_m)} \big( \cosh (\Omega t/(2\tau_m))+ \frac{1}{\Omega} \sinh (\Omega t/(2\tau_m)) \big) \cos (kz),
\label{NS_2}
\end{equation} 
 where 
\begin{equation}
\Omega = \sqrt{1-4 \tau_m \frac{\eta}{\rho} k^2}.
\end{equation}
For large $k$ equation (\ref{NS_2}) leads to a similar solution as equation (\ref{NS}). The viscosity from this method is given by \citep{palmer:94:0} 
\begin{equation}
\eta (k)= \eta(0)(1-ak^2)+O(k^4). 
\label{eta_TCAF}
\end{equation} 

\subsubsection{The periodic perturbation method }
As mentioned earlier, in PP method an external force is applied to the system. The external field leads to development of a velocity field $\bold {u}$  throughout the system according to the Navier-Stokes equation.
For a force only in the $x$ direction, the applied acceleration $a_x$ for a periodic system is given as
\begin{equation}
a_x(z)= \alpha  \cos(kz); \hspace{2cm} k=\frac{2 \pi}{l_z},
\end{equation}
where, $l_z$ is the height of the box and $\alpha$ is the amplitude of the acceleration. With the initial value of $u_z(x)=0$, the resulting  velocity profile has amplitude  
\begin{equation}
\nu (1-e^{-t/\tau_r}),
\end{equation}
where 
\begin{equation}
\nu=\alpha \frac{\rho}{\eta k^2}.
\label{nu}
\end{equation}
Thus, at each time step the average velocity can be measured and viscosity can be obtained from Eq.~\ref{nu}.
In order to obtain accurate results efficiently, the parameters of the periodic external force must be chosen carefully. 
If the velocity profile does not have large fluctuations, less statistics needs to be collected, and thus the simulation time is shorter.
This can be achieved with a large amplitude.
However, the shear rate should also not be so high that the system moves too far from equilibrium. For more details about the method and estimation of parameters see ref.~\cite{hess:02:0}.

\section{Shear viscosity from simulations} 
\begin{figure}
\centering
\vspace{1cm} 
\includegraphics[scale=0.34]{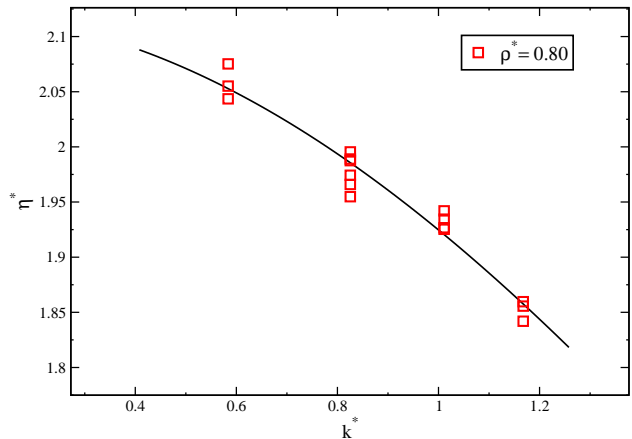}
\vspace{-0.7cm}
\caption{{   Shear viscosity of PHS obtained from simulation at   $\rho^* =0.80$ (data). The solid line shows  the fit to Eq.~\ref{eta_TCAF} (TCAF method). $k^* =k\sigma$. }}
\label{tcaf_example}
\vspace{1cm}
\end{figure}

\begin{figure}
\centering
\includegraphics[scale=0.9]{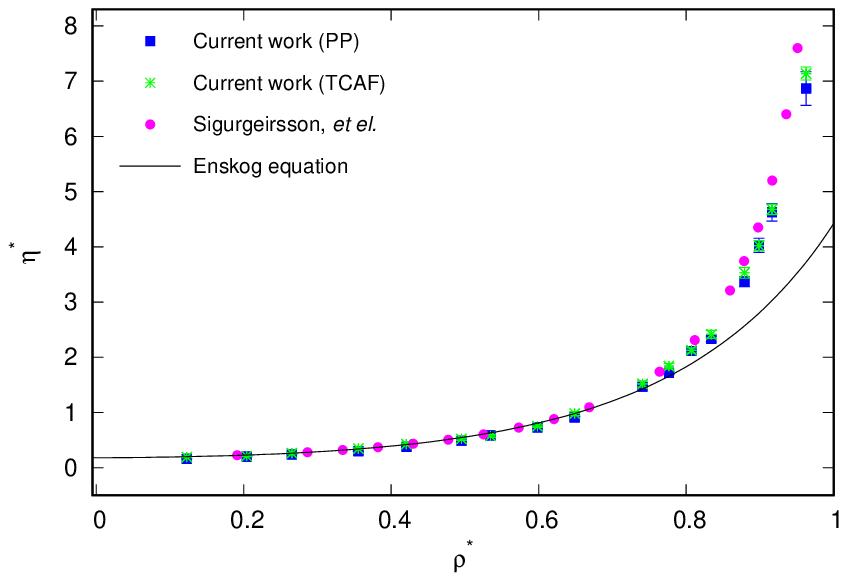}
\includegraphics[scale=0.9]{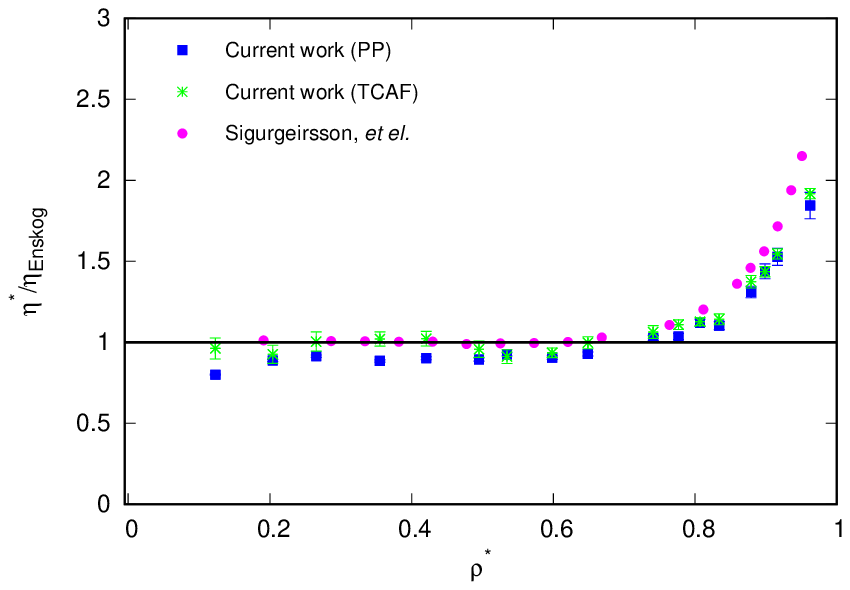}
\caption{ (Left) Shear viscosity of PHS model    from TCAF method (crosses) and from PP method (squares). The red circles are results of Sigurgeirsson {\it {et al.}} \cite{heyes:09:0} and the  solid line is the Enskog expression, Eq.~\ref{Enskog}. {  (Right) The same as the left plot with the Enskog equation as  a base function. }  }
\label{vis_TCAF}
\end{figure}

We first discuss  the TCAF method.
We take the final configurations of the NPT simulations explained in previous section and  simulate the system in NVT ensemble for $t^*= 3000$ with $\delta t^*= 0.0011$ time step. In the TCAF method the correct values of viscosity are obtained when the velocity profile is not coupled to a heat bath. Therefore, a  Berendsen thermostat with a long coupling time of $t^*=11$ is used in order to minimize the influence of the thermostat, see details in ref.~\cite{hess:02:0}. The velocity and coordinates are stored every $\delta t^*= 0.011$.  The neighbor list was updated every $\delta t^*= 0.005$.
 The resulting k-dependent 
viscosities from the simulations  are fitted 
to  Eq.~\ref{eta_TCAF}  to obtain the viscosities. 
Fig.~\ref{tcaf_example} shows an example of  simulation  result  at density  $\rho^*=0.8$. The resulting viscosity from the fit gives  $\eta^*=2.11 \pm .043 $ in reduced unit, $\eta^*=\eta (\frac{\sigma ^2}{\sqrt {mk_B T} })$. 
The  obtained values of $\eta^*  $ from this method   are given in Table.~\ref{vis_PHS} for various pressures/densities (the last two columns), and are shown in   Figure.~\ref{vis_TCAF} as  crosses.  {   The errors in the given  viscosities from this method are obtained  from block averaging over k-dependent
viscosities  plus uncertainty  of  the fittings to Eq.~\ref{eta_TCAF}.  }

For the PP method, we simulate the system in  NVT ensemble for $t^*= 3000$ with an  addition of  an externally imposed acceleration  \cite{hess:02:0}. The coupling time of the Berendsen thermostat is set to $t^*= 2.84$ in oder to remove the energy introduced in the system by the external force more rapidly. 
The optimal acceleration amplitude was estimated (see Eqs. 22, 26, 27 in ref.~\cite{hess:02:0}) to be around $  \alpha^*= 0.054$, which is big enough to develop sufficient shear, but also not so big that the system moves too far from equilibrium. 
{  One disadvantage of this method  is that the obtained viscosities depend  on the chosen amplitude  $\alpha$ \cite{hess:02:0},  moreover, the chosen amplitude should be changed with the density in the systems. }
 We have started the analysis at time $300$ reduced units after the start of applying the force, so that there is enough time to develop a steady shear amplitude.
The obtained values of $\eta^*  $ from this method are given  in Table.~\ref{vis_PHS} (the fifth and sixth  columns).   The errors in the viscosities are  obtained from block averaging.
The results are included in Figure.~\ref{vis_TCAF}, along with the results of the TCAF methods and the MD results of the true HS model reported by Sigurgeirsson  {\it {et al.}} \cite{heyes:09:0}. We give the simulation results for $\rho^*>0.1$, because when the system is dilute the mean-free-path of the particles becomes large and  the  simulation box should be large enough to make the collisions to occur enough.  That is computationally expensive. \\
The results from to the TCAF method show better agreement than the PP method. The reason is that the chosen amplitude  $\alpha$ has  effect on the viscosity, as studied comprehensively in ref.~\cite{hess:02:0} and in order to obtain more accurate estimations from this method one should try several amplitudes for each density, which is time-consuming.
\begin{table}
\center
\vspace{1cm}
\scalebox{1}{
\begin{tabular}{|c |c  |c  |c  |c  |c   |c   |c  |}
\hline
 $P^*$ &  error in $P^*$ &   $\rho ^*$   & error in   $\rho ^*$    &   $\eta ^*$ (PP)  &  error in  $\eta ^*$ (PP)  & $\eta ^*$  (TCAF)   &error in  $\eta ^*$  (TCAF)    \\
        \hline
0.15978 & 0.00069 & 0.12286 & 0.00014 & 0.19115 & 0.01286 & 0.15894 & 0.00080 \\ \hline
0.31944 & 0.00182 & 0.20395 & 0.00028 & 0.21134 & 0.01275 & 0.20263 & 0.00158 \\ \hline
0.47930 & 0.00284 & 0.26531 & 0.00025 & 0.26418 & 0.01572 & 0.24035 & 0.00203 \\ \hline
0.79872 & 0.00378 & 0.35532 & 0.00026 & 0.34591 & 0.01497 & 0.30003 & 0.00234 \\ \hline
1.11829 & 0.00517 & 0.42067 & 0.00035 & 0.42858 & 0.01868 & 0.37814 & 0.00368 \\ \hline
1.59719 & 0.00750 & 0.49526 & 0.00029 & 0.52196 & 0.02760 & 0.48749 & 0.00539 \\ \hline
1.91703 & 0.00988 & 0.53465 & 0.00030 & 0.57531 & 0.02666 & 0.58406 & 0.00673 \\ \hline
2.55643 & 0.01294 & 0.59829 & 0.00032 & 0.75286 & 0.02338 & 0.72918 & 0.00862 \\ \hline
3.19444 & 0.01563 & 0.64875 & 0.00028 & 0.97986 & 0.03616 & 0.91150 & 0.01233 \\ \hline
4.79239 & 0.02033 & 0.74088 & 0.00033 & 1.51894 & 0.05567 & 1.46623 & 0.02304 \\ \hline
5.59744 & 0.03030 & 0.77666 & 0.00164 & 1.83790 & 0.04855 & 1.71697 & 0.03116 \\ \hline
6.38936 & 0.02893 & 0.80720 & 0.00027 & 2.12458 & 0.04511 & 2.11137 & 0.04336 \\ \hline
7.18685 & 0.02710 & 0.83400 & 0.00021 & 2.41335 & 0.07214 & 2.32994 & 0.05179 \\ \hline
8.78463 & 0.03660 & 0.87911 & 0.00034 & 3.52977 & 0.09911 & 3.36685 & 0.08977 \\ \hline
9.58201 & 0.04701 & 0.89880 & 0.00030 & 4.02040 & 0.09138 & 4.02834 & 0.12877 \\ \hline
10.37838 & 0.04053 & 0.91643 & 0.00026 & 4.67897 & 0.09505 & 4.62323 & 0.15965 \\ \hline
12.77719 & 0.04276 & 0.96250 & 0.00020 & 7.13797 & 0.11703 & 6.86586 & 0.30399 \\ \hline
\end{tabular}}
\vspace{0.2cm}
\caption{Shear viscosity obtained from the simulations from the two different methods, PP and TCAF. The given shear viscosities are dimensionless; $\eta^*=\eta (\frac{\sigma ^2}{\sqrt {mk_B T} })$. }
\label{vis_PHS}
\end{table}
\section{Comparison with theory}
The Enskog expression for the viscosity of a fluid of hard spheres is  \cite{Chapman:52:0,Santos:16:0,Viswanath:07:0,Sengers:00:0,Lucas:79:0}
\begin{equation}
\eta= \eta_0 \bigg [ g^{-1}(\sigma)+0.8 ~V_{excl}~ \rho + 0.776~ V^2_{excl} ~\rho^2 ~g(\sigma)\bigg],
\label{Enskog}
\end{equation} 
where 
\begin{equation}
\hspace{2cm} \eta_0=\frac{5}{16\sigma^2}\sqrt{\frac{mk_B T}{\pi}},
\label{ensk}
\end{equation} 
and $V_{excl} $ is the excluded volume of HS, $V_{excl}=\frac{2\pi\sigma^3}{3}$,  and $g(\sigma)$ is the  radial distribution function (rdf) at contact.  $\eta_0$ is the  zero-density viscosity.
The rdf at contact can be obtained directly from the Carnahan-Starling equation of state, which yields,
   \begin{equation}
g(\sigma) = \frac{1-\xi/2}{(1-\xi)^3},
\label{CS}
\end{equation} 
where $ \xi=\frac{\pi \rho }{6}$ is the volume fraction.
It can also be found from the Percus-Yevick  approximation,~\cite{yelash:01:00,yevick:58:00}.
   \begin{equation}
g(\sigma) = \frac{1+\xi/2}{(1-\xi)^2}.
\label{PY}
\end{equation} 
The latter  is more accurate at high density metastable fluid region \cite{heyes:09:0}.

In Figure.~\ref{vis_TCAF} we compare the simulation results to the Enskog theory  and to the  previous MD simulations of  true HS simulations by Sigurgeirsson  {\it {et al.}}~\cite{heyes:09:0}.  The figures present a qualitative agreement with both. Similar to the ref.~\cite{heyes:09:0}  the Enskog theory produces good agreements only for low- to mid-density ranges and fails at high densities, since it does not take into account correlated collisions.
That is the reason for deviations at high densities in Fig.~\ref{vis_TCAF}  \cite{Santos:16:0,Viswanath:07:0,Sengers:00:0,Lucas:79:0}. 
\begin{figure}[H]
\centering
\vspace{1cm}
\includegraphics[scale=0.33]{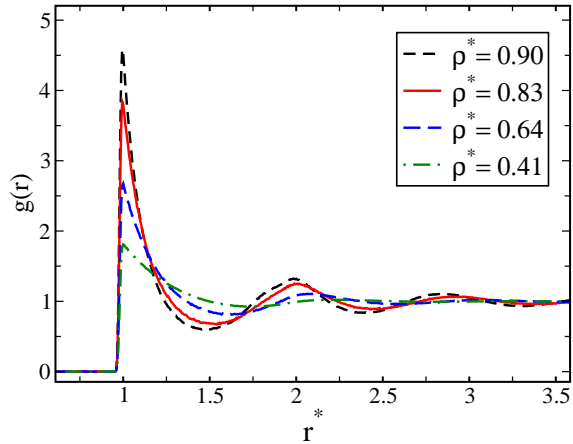}
\vspace{-0.5cm}
\caption{Radial distribution functions of PHS obtained from the simulations at several densities.  }
\label{rdf_ar}
\end{figure}
\begin{figure}[H]
\centering
\vspace{1cm}
\includegraphics[scale=0.33]{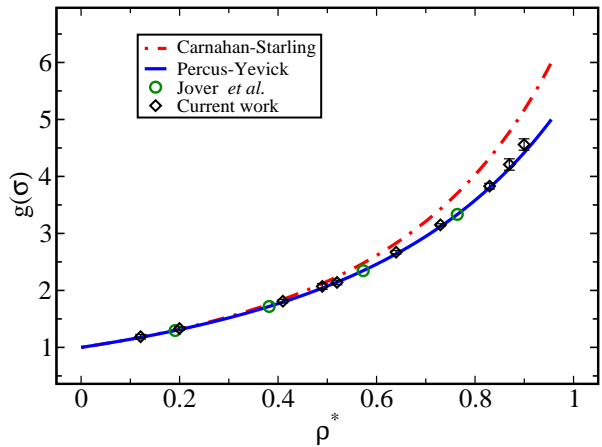}
\vspace{-0.5cm}
\caption{Radial distribution functions of PHS at contact obtained from the simulations (diamonds) along  with  the theoretical expressions of Carnahan-Starling  (dot-dashed line) and Percus-Yevick (solid line). The circles are MD simulations by  Jover {\it {et al.}}.}
\label{rdf_contact}
\end{figure}

We also compare the radial distribution function at contact obtained from simulations (only for several densities) to the theoretical prediction from using the Carnahan-Starling equation of state and the Percus-Yevick approximation.
In order to obtain the contact values of the radial distribution functions from simulations, we first obtain the rdf profiles for each  density, see Figure.~\ref{rdf_ar} (the results are taken from the  simulations explained in Section.~\ref{sec3}). The contact values then are collected at $r^*=1$. The values of the  rdf at contacts are given in black diamonds in Fig.~\ref{rdf_contact}. Simulation results from ref.~\cite{jover:12:0} are also included as circles.
Eqs.~\ref{CS}-\ref{PY} are represented in this figure by dot-dashed and solid  lines, respectively. As seen in the figure, the simulation results agree well with the Percus-Yevick expression and with results produced  by  Jover {\it {et al.}} \cite{jover:12:0}.
As expected Carnahan-Starling's expression deviates at higher densities, because it does not accurately capture higher virial coefficients.
\section{Conclusions}
We have tested the reliability of the  pseudo hard sphere (PHS) potential for calculating the shear viscosity of hard spheres. We have used molecular-dynamics simulations (with GROMACS) for this purpose and obtained shear viscosities of PHS from two different methods; transverse-current autocorrelation function and the periodic perturbation method.    

We have run simulations under the same conditions that have previously been confirmed to
 represents the hard-sphere equilibrium properties as demonstrated by  Jover {\it {et al.}} \cite{jover:12:0}.
We have compared the shear viscosities from simulations to the Enskog theory and to available MD simulation data  ~\cite{heyes:09:0}.  Similar to  ~\cite{heyes:09:0}, the comparison with Enskog equation shows a good agreement for densities up to $\rho\sigma^3\le 0.65$.
In addition, contact values of the radial distribution functions agree with both Carnahan-Starling (at low densities)  and Percus-Yevick expression (up to $\rho\sigma^3\approx0.85$).

The validation of shear viscosity of PHS model helps to use it directly into the state of art simulations packages like LAMMPS, GROMACS or  NAMD  (which provide high-speed simulations) in order to study  complex and large systems.
This in turn greatly simplifies its use to support development of analytical theory for similar models based on hard spheres, such as dipolar hard spheres, which will aid the development of transport theory  as well as empirical approaches to more complex liquids.
\section{Acknowledgments}
The work has been supported by National Infrastructure for Computational Science in Norway
(UNINETT Sigma2) with computer timed for the Center for High Performance Computing (NN9573K and NN9572K). The authors acknowledge The Research Council of Norway for NFR project number 275507  and The Faculty of Engineering, Norwegian University of Science and Technology (NTNU), for financial support. FP acknowledges  Prof. Erich  M{\"u}ller for providing data for radial distribution of PHS. 
\bibliography{ref} 
\bibliographystyle{ieeetr}

\end{document}